\RequirePackage{snapshot}
\documentclass[
reprint,
superscriptaddress,
 amsmath,amssymb,
 aps,
]{revtex4-1}

\usepackage{graphicx}
\usepackage{dcolumn}
\usepackage{bm}
\usepackage{color,amsmath}
\usepackage{physics}
\usepackage{subfiles}

\usepackage[normalem]{ulem}
\usepackage[usenames,dvipsnames]{xcolor}
\usepackage{hyperref}
\hypersetup{
    colorlinks,
    linkcolor={violet},
    citecolor={ForestGreen},
    urlcolor={teal}
}

\usepackage{subfigure}

\newcommand{\new}[1]{\textcolor{black}{#1}}

\begin{document}

\title{Ergodicity Breaking and Deviation from Eigenstate Thermalisation \\ in Relativistic Quantum Field Theory} 

\author{{Miha Srdinsek}}
\email{miha.srdinsek@upmc.fr}
\affiliation{Institut des Sciences du Calcul et des Données (ISCD), Sorbonne Université, 4 Place Jussieu, 75005 Paris, France}
\affiliation{Institut de Minéralogie, de Physique des Matériaux et de Cosmochimie (IMPMC), Sorbonne Université, CNRS UMR 7590, MNHM, 4 Place Jussieu, 75005 Paris, France}
\affiliation{Processus d’Activation Sélectif par Transfert d’Energie Uni-électronique ou Radiative (PASTEUR), CNRS UMR 8640, Département de Chimie, École Normale Superieure, 24 rue Lhomond, 75005 Paris, France
}
\author{{Tomaz Prosen}}
\affiliation{Faculty of Mathematics and Physics,
University of Ljubljana, Jadranska 19, SI-1000 Ljubljana, Slovenia}
\author{{Spyros Sotiriadis}}
\affiliation{Institute of Theoretical and Computational Physics,
Department of Physics, 
University of Crete, 71003 Heraklion, Greece}
\affiliation{Dahlem Center for Complex Quantum Systems, Freie Universit{\"a}t Berlin, 14195 Berlin, Germany}

\date{\today}

\begin{abstract}
The validity of the ergodic hypothesis in quantum systems can be rephrased in the form of the eigenstate thermalisation hypothesis (ETH), a set of statistical properties for the matrix elements of local observables in energy eigenstates, which is expected to hold in any ergodic system. We test the ETH in a nonintegrable model of relativistic quantum field theory (QFT) using the numerical method of Hamiltonian truncation in combination with analytical arguments based on Lorentz symmetry and renormalisation group theory. We find that there is an infinite sequence of eigenstates with the characteristics of quantum many-body scars, that is, exceptional eigenstates with observable expectation values that lie far from thermal values, and we show that these states are one-quasiparticle states. We argue that in the thermodynamic limit the eigenstates cover the entire area between two diverging lines: the line of one-quasiparticle states, whose direction is dictated by relativistic kinematics, and the thermal average line. Our results suggest that the strong version of the ETH is violated in any relativistic QFT whose spectrum admits a quasiparticle description.
\end{abstract}

\maketitle  

\emph{Introduction} -- 
One of the main theoretical concepts for understanding and testing ergodicity in quantum systems is the eigenstate thermalisation hypothesis (ETH) \cite{Deutsch91, Srednicki94}, which states that in a chaotic quantum system -- to be termed ergodic -- the expectation values of a local observable in generic high-energy eigenstates concentrate at the corresponding thermal expectation values as we approach the thermodynamic limit. The validity of the ETH in quantum many-body systems has been studied and verified numerically in a large number of nonintegrable lattice models \cite{Rigol08, Reimann2015,D_Alessio_2016, Deutsch2018, NandkishoreHuse2015, Santos-Rigol,Moessner14, D_Alessio_2016,Borgonovi, Foini-Kurchan}, 
even though exceptions to this rule do exist, at least in a weak sense. For instance, in models exhibiting so-called \emph{quantum many-body scars} (QMBS) \cite{Turner18a,Turner18b,Lin18,Choi19,Ho19,Bull19a,Bull19b,Schecter19,Michailidis20revivals, Moudgalya2018, Moudgalya2018b, Iadecola2020, Mark2020, Chattopadhyay2020,O'Dea2020,Pakrouski2020,Ren2021,Langlett2022,QMBS_perspective}, expectation values of local observables are anomalously distant from the corresponding thermal values for an infinite sequence -- yet of measure zero -- of eigenstates. Whenever such atypical states are present, the crucial question that arises is what happens in the thermodynamic limit: Do these states vanish (strong ETH), do they persist but correspond to a zero measure subset of the spectrum (weak ETH), or do they correspond to a finite density? Only the strong version of the ETH can guarantee thermalisation \cite{Biroli10}.

A class of quantum many-body models that remains largely unexplored in this respect is that of nonintegrable quantum field theories (QFT), i.e., relativistic models of quantum fields,  which constitute the theoretical basis of particle physics and adjacent research areas. QFT models are defined over Hilbert spaces of infinite dimension, for which reason the numerical study of their spectra is inevitably approximate and remains an especially challenging task. Some pioneering studies of the validity of the ETH in QFT have focused on conformal field theories (CFTs) \cite{Cardy2014,Pallab2017, Datta2019, Song2017, Lashkari_2018, Dymarsky2019, Kanato2022,Zhiyuan2022}, which exhibit quantum chaotic characteristics in the limit of large central charge \cite{Liam2015, Curtis2015,Tarek2016, Asplund2015, Turiaci2016, Kusuki2019, Kudler-Flam2020, Kudler-Flam2021}. In a general nonintegrable QFT, numerical tests of the ETH can be performed using lattice discretisation \cite{scars_Schwinger_short, scars_Schwinger} or Hamiltonian truncation (HT) methods \cite{Konik_rare_states, Srdinsek_2020, Delacretaz2023}. Studying a class of prototypical nonintegrable models of $(1+1)$D QFT based on HT, we have recently demonstrated that, although their level spacing statistics are consistent with random matrix theory predictions, the statistics of their eigenvector components is markedly different from that of random matrices \cite{Srdinsek2021}. {The observation was further confirmed by Delacrétaz \textit{et al.} \cite{Delacretaz2023} for the case of the $\phi^4$ model.} Eigenvector statistics is strongly related to the validity of the ETH; therefore our observation naturally raises the question of whether the ETH is satisfied in these models.

In this Letter, we argue that, unlike in typical lattice models, violations of the ETH are commonplace in a large class of nonintegrable QFTs. First, there are exceptional eigenstates with QMBS characteristics. Similar special states have been observed also in other QFT models \cite{Konik_rare_states, Delacretaz2023} and in one of them \cite{Konik_rare_states} their presence has been associated to confinement, resulting in the absence of thermalisation after a quench \cite{Kormos17}. Second, the eigenstates do not all concentrate close to a line in the thermodynamic limit, as prescribed by the ETH, but remain spread in a wide area, from an edge line where QMBS-like states are located to the thermal average line.

Our numerical results, based on HT applied to a nonintegrable QFT, demonstrate the existence of eigenstates that have QMBS characteristics \cite{Turner18a,Turner18b}: They are outliers in ETH diagrams, span the entire energy range of the spectrum, are approximately equidistant in energy and have low (Fourier-space) entanglement. In addition, the presence of such states turns out to persist as the system size or interaction strength increases, at least within the limits of applicability of our method. 

One interpretation of QMBS is in terms of quasiparticle states at unusually high-energy scales in the spectrum \cite{QMBS_perspective}. We find that the scar-like states we observe can indeed be explained as quasiparticle states. Based on this interpretation, we show that their presence in the thermodynamic limit, at high energies and for arbitrary interaction strength follows from relativistic kinematics, which at the same time means that they correspond to physically relevant quantum states. 

On the other hand, based on renormalization group (RG) theory it is expected that in the high-energy (ultraviolet) limit the thermal average line follows that of the CFT that describes the ultraviolet fixed point of the given model. This is because in this limit any RG relevant operator by definition reduces to only a weak perturbation of the CFT \cite{Konik-Adamov, Hogervorst2014}. The asymptotic CFT thermal line generally diverges from the line corresponding to one-quasiparticle states, which means that eigenstates are expected to cover the entire area between them. Based on the above arguments, we conclude that deviations from ETH are expected in any Lorentz invariant relativistic QFT model whose spectrum admits a quasiparticle description.
 
\emph{Model and method --} 
We consider the \emph{double sine-Gordon} (DSG) model \cite{Delfino-Mussardo_DSG,Bajnok2000,Bajnok01,Toth2004,Takacs_DSG,Mussardo_DSG,Roy-Lukyanov}, a nonintegrable $(1+1)$D bosonic QFT that is described by the Hamiltonian {(in units $\hbar=c=1$)} 
\begin{align}
	H_\mathrm{DSG} & = H_{0} - \lambda_1 V(\beta_1) - \lambda_2 V(\beta_2), \label{H_DSG} \\
	H_{0} & = \int_{0}^{L} :\! \left(\tfrac12 \Pi^{2} + \tfrac12 \left(\partial_{x}  \phi\right)^{2} \right) \!: \; \mathrm{d} x , \label{H0} \\  
	V(\beta) & = \int_{0}^{L} :\! \cos \beta \phi \!: \mathrm{d} x . \label{V} 
\end{align}
The low-energy spectrum of DSG can be studied efficiently using HT in the basis of the free massless boson model $H_0$ truncated by imposing a maximum energy cutoff \cite{Takacs_DSG,Toth2004,Srdinsek_2020}. This is an application of the truncated conformal space approach (TCSA) \cite{Yurov-Zamo,Yurov-Zamo_2}, where the unperturbed model is a CFT and the construction of the perturbation matrix is facilitated using the algebraic CFT toolkit. TCSA has been used to study spectral properties, as well as equilibrium and quench dynamics in the sine-Gordon model and perturbations thereof \cite{KST,Kukuljan_2020,Kukuljan_Schwinger,TCSA_entanglement_entropy,Srdinsek2021,Horvath2022_CFTCSA,Horvath2022}. The convergence of the numerically computed eigendecomposition for increasing cutoff is guaranteed for perturbations that are RG relevant operators and is faster when the scaling dimension of the perturbation is smaller, meaning that TCSA eigenstates can be used for testing the ETH \footnote{In DSG, convergence is relatively fast compared to polynomial interactions (e.g. the $\phi^4$ model) because the scaling dimension of the cosine operator, which is $\Delta=\beta^2/(4\pi)$, can be chosen arbitrarily small.}. The smaller the system size $L$, the interaction strength parameters $\lambda_i$, and frequencies $\beta_i$ in Eq.~\ref{V}, the faster the convergence.
\begin{figure*}[th]
    \centering
        \includegraphics[width = 1.0\linewidth]{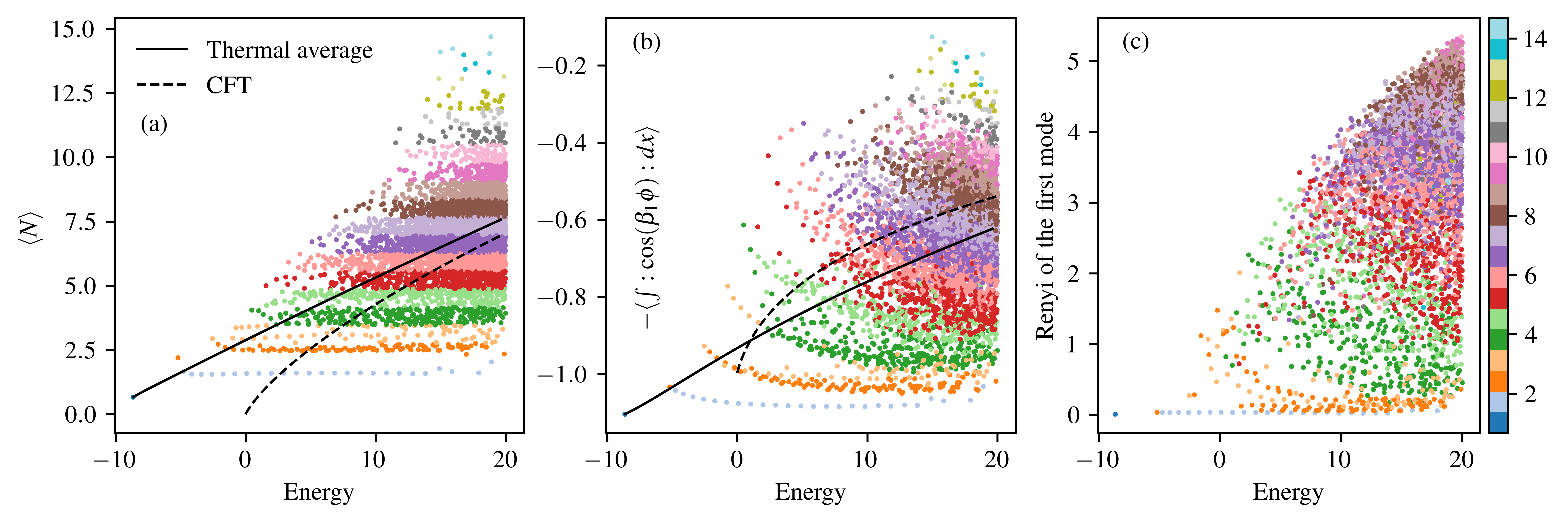}
    \caption{{Eigenstate expectation values of two different observables ((a) particle number $N$ and (b) the cosine potential term $V(\beta_1)$) and (c) of the R\'enyi entanglement entropy versus the corresponding energy eigenvalues in the units of $\pi/L$, for DSG at $\mu=0.5$ (where $\mu\equiv\lambda (L/\pi)^2$). The colours indicate the expectation value of particle number $N$. The black lines are numerically computed thermal curves of expectation values in thermal states of $H_\text{DSG}$ (full) or $H_{0}$ (dashed) versus the corresponding mean thermal energy. These lines are expected to become asymptotically parallel at high energies.
    }
    \label{fig:1}}
\end{figure*}
\begin{figure*}
    \centering
        \includegraphics[width = 1.0\linewidth]{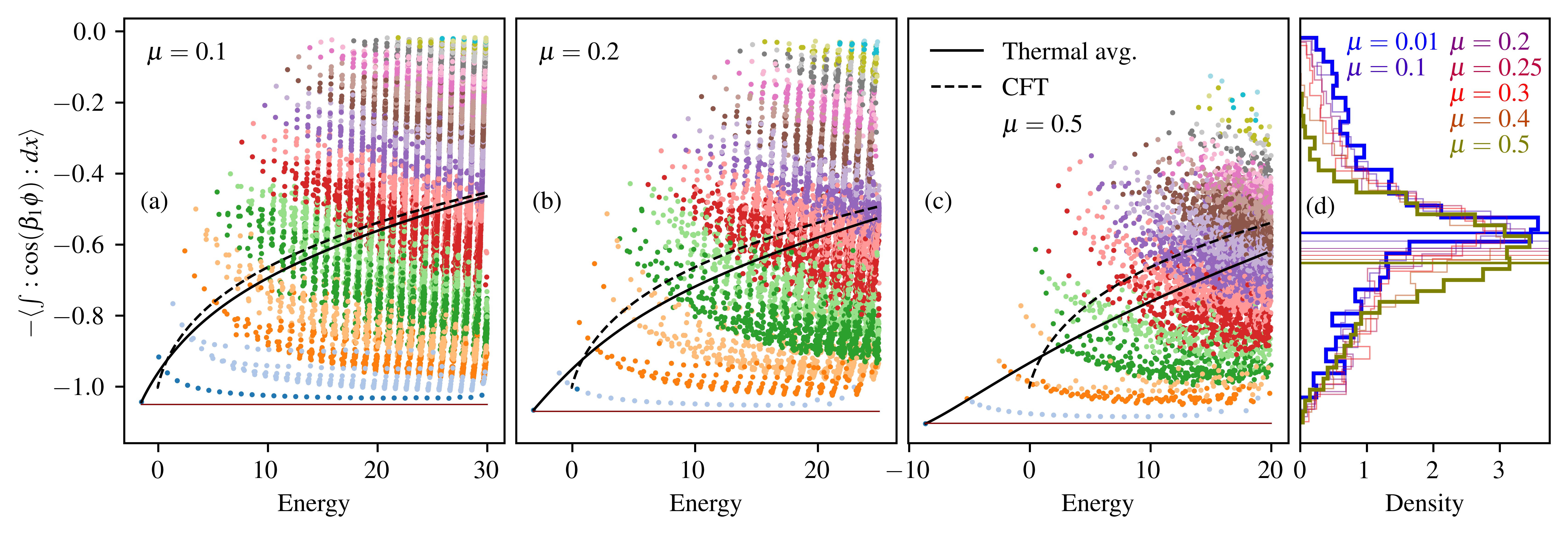}
    \caption{{{(a-c)} The same as in Fig.~\ref{fig:1} for a single observable, {$V(\beta_1)$}, at increasing values of the perturbation strength $\mu\equiv\lambda (L/\pi)^2$. At $\mu=0.1,0.2,$ and $0.5$ there are $28871$, $10803$, and $5483$ converged states,  respectively. {(d) Histogram of eigenstate expectation values of $V(\beta_1)$ in the energy window $E=19\pm2$ at various $\mu$.}}
    \label{fig:2}}
\end{figure*}

\emph{Numerical results} -- We study the spectrum of DSG for Dirichlet boundary conditions at $\beta_1 = 1$ and $\beta_2 = 2.5$ with equal mixing parameters $\lambda_1 = \lambda_2=\lambda$, in which case the model is far from all integrable regimes. In particular, its level spacing statistics are consistent with RMT predictions for any value of the interaction strength $\lambda$ even in proximity to the CFT point \cite{Srdinsek2021}. At the same time, the spectrum converges up to sufficiently high energies and for interactions sufficiently far from the perturbative regime. The system size dependence of the operators $H_0$ and $V(\beta)$ {in the free massless boson basis} can be shown to be simply $H_0\propto 1/L$ and $V(\beta)\propto L$, respectively, so it is convenient to compute the spectrum of the rescaled Hamiltonian $\tilde{H}_\mathrm{DSG} = \tilde{H}_{0} - \lambda (L/\pi)^2[\tilde{V}(\beta_1)+\tilde{V}(\beta_2)]$ where $\tilde{H}_{0}={H}_{0}(L/\pi)$, whose energy eigenvalues are integers, and $\tilde{V}(\beta)={V}(\beta)(\pi/L)$. This way it is clear that there is only one free parameter, $\mu\equiv\lambda (L/\pi)^2$, and increasing the interaction coupling has essentially the same effect as increasing the system size. 

As observables, we use the number operator $N$ in the CFT basis and the potential energy operators $V(\beta_i)$. Note that $N$ is not a conserved quantity in the DSG. These operators are integrals of local densities, therefore having extensive expectation values in ground or equilibrium states, which makes them suitable for testing the ETH. Given that the unperturbed CFT model is diagonal in terms of Fourier modes, we can compute, as a measure of entanglement in the eigenstates, the R\'enyi entanglement entropy of one Fourier mode with respect to the rest. In lattice models, the entanglement entropy of an eigenstate computed in the local basis encodes the total correlation between the subsystem in focus and its complement, with localised eigenstates showing low entanglement compared to delocalised ones. Similarly, Fourier space entanglement is low for eigenstates that are `localised' in the Fourier space.

Fig.~\ref{fig:1} shows numerical results for the expectation values of these quantities in the eigenstates versus their energy. We set $\mu=0.5$, which corresponds to a good trade-off between remoteness from the perturbative limit and accuracy. While the spectrum exhibits chaotic level spacing statistics, the scatter plots look qualitatively very different from those of chaotic models. In contrast to what the ETH prescribes, the eigenstates span a wide area rather than being concentrated close to the thermal curve, and they are organised in separated families of states, a structure reminiscent of integrable models. In particular, the states with the lowest particle number $N$ form a special family of points located at the edge of the spectra, well aligned along smooth curves spanning the entire energy range, and having almost equidistant energies. These states have very low Fourier space entanglement entropy, which is characteristic of QMBS.

Varying $\mu$ from 0 to 1 we observe that some but, as we will see, not all characteristics of this structure are inherited from the perturbative limit. As shown in Fig.~\ref{fig:2}, in this limit the spectrum is split into vertically aligned and equidistant families of points, and the lowest $N$ eigenstates that look like scars are located at the edge of each family. Given that the unperturbed model is a CFT, its spectrum consists of degenerate energy sectors at equidistant energies with degeneracy numbers increasing rapidly with the energy. A weak perturbation of such a highly degenerate model mixes states predominantly within the same degenerate sector, resulting in decoupled energy spectra. It is, therefore, not surprising that states at the edges of these spectra are less random than in the middle \cite{SM}. The above argument provides a generic justification for the emergence of scar-like states in any model that is a weak perturbation of a highly degenerate model with equidistant spectrum. It cannot, however, explain the presence of scars away from the perturbative limit.

Our numerical results indicate that, in our model, the scar states we observe do not disappear upon increasing $\mu$, which is effectively the same as approaching the thermodynamic limit. Indeed, at $\mu\sim 0.5$, the mixing of the originally decoupled sectors is strong enough that the spectrum is no longer split into gap-separated sectors; nevertheless, scars are still present and aligned along curves that deviate strongly from the thermal average, showing no tendency to approach it (Fig.~\ref{fig:2}). As shown in Fig.~\ref{fig:2}.d, focusing on a fixed energy window the density of states close to the bottom edge does not decrease for increasing $\mu$, unlike that of the upper edge, which contracts closer to the thermal average. While we cannot obtain a definite answer to what happens at stronger interaction by means of numerical HT methods, a useful insight into the nature of these states can be gained based on analytical arguments discussed below.

\emph{The role of Lorentz symmetry} -- Quantum scars may be interpreted as quasiparticle states appearing at unusually high energies. In a general interacting model, low-energy eigenstates can be described as quasiparticles, i.e., collective excitations formed by many real particles moving coordinately under the effect of their mutual interactions. However, at higher energies quasiparticles typically become unstable due to the large number of possibilities to decay to other quasiparticles, which is why the existence of stable-quasiparticle states at higher energies is surprising.

{The spectrum of a typical QFT model admits a quasiparticle description.} The ground state is the state that is empty of quasiparticles, and the first excited state contains only one quasiparticle at rest (the lightest of all, if many) with other low-energy states corresponding to moving quasiparticles or more than one quasiparticle scattering with each other. In massive QFT models, eigenstates with finite energy correspond to a finite number of quasiparticles, as there is an energy threshold for the production of massive quasiparticles. For this reason any eigenstate of a massive QFT can be described by special relativity kinematics for the set of constituent massive quasiparticles. 

As a result, signatures of Lorentz invariance are clearly manifested in the structure of QFT spectra on the $(P,E)$ plane, where $P$ and $E$ are the total momentum and energy expectation values, respectively, of the eigenstates. Indeed, for a massive QFT in infinite space with periodic boundary conditions, all excited states are enveloped by the hyperbola $E=\sqrt{P^{2}+M^{2}}$, where the ground state energy is set to zero. In finite-size systems, the spectra are discrete rather than continuous but otherwise exhibit the same characteristics, subject to two types of decaying corrections compared to the infinite-size case \cite{Luscher1986i,Luscher1986ii}. Importantly, the above described qualitative structure of the spectrum is valid independently of the interaction strength. Nonperturbative effects at strong interactions are encoded in the complex nature of the quasiparticles (e.g. parameters like mass or charge, scattering amplitudes, etc., depend nontrivially on the interaction), whereas qualitative characteristics of the spectra are determined by relativistic kinematical constraints \cite{Eden_Analytic_S-matrix}. 
\begin{figure}
    \centering
\includegraphics[width = \linewidth]{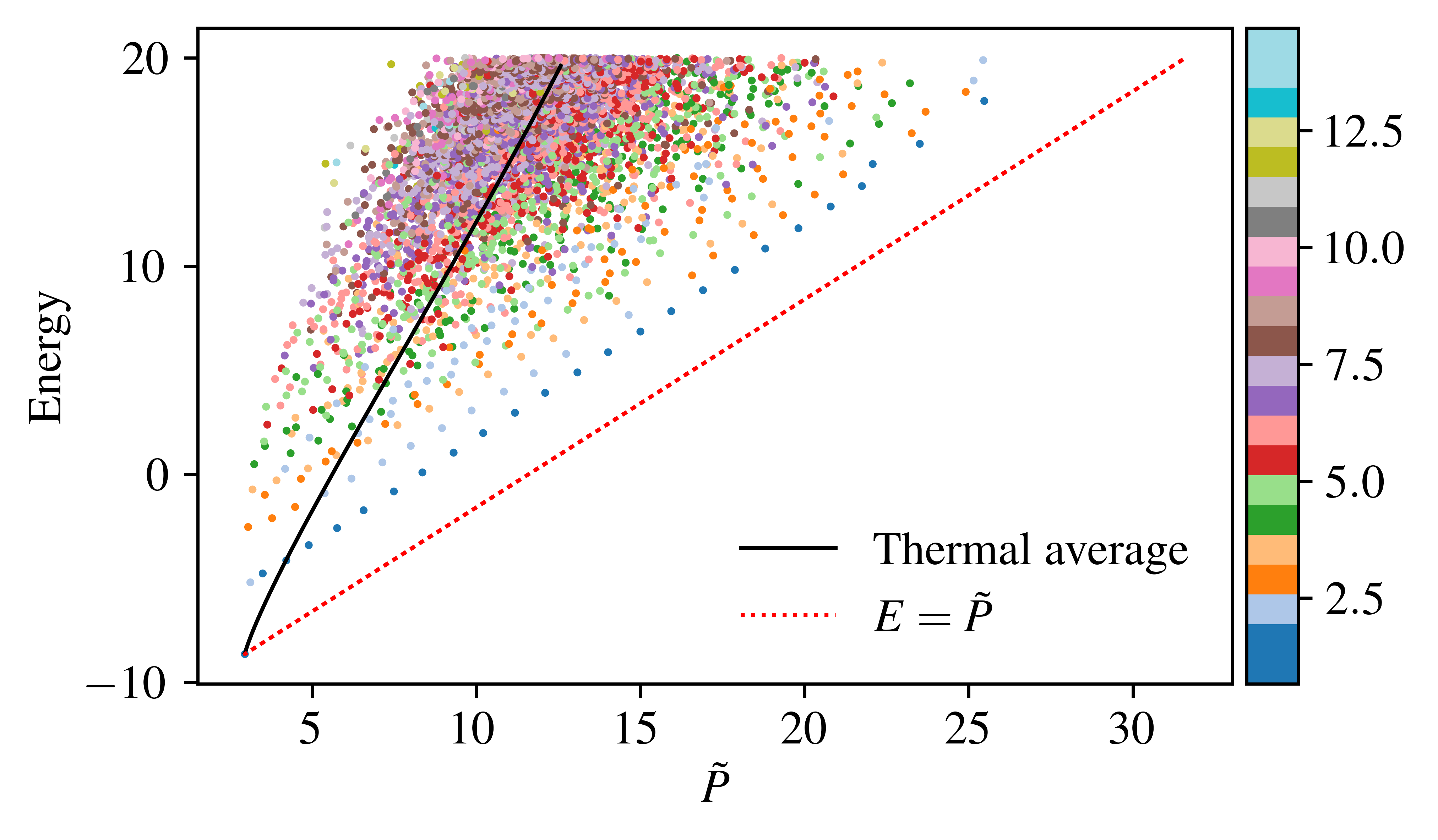}
    \caption{{
    Scatter plot of $ \tilde{P}=\sqrt{\langle P^2 \rangle}$ in each eigenstate versus the corresponding energy eigenvalue $E$ in units of $\pi/L$ for the numerically computed DSG spectra at $\mu=0.5$. The colours indicate the particle number expectation value. \label{fig:E-P}
  }}
\end{figure}
\begin{figure}
\centering
        \includegraphics[width = 1\linewidth]{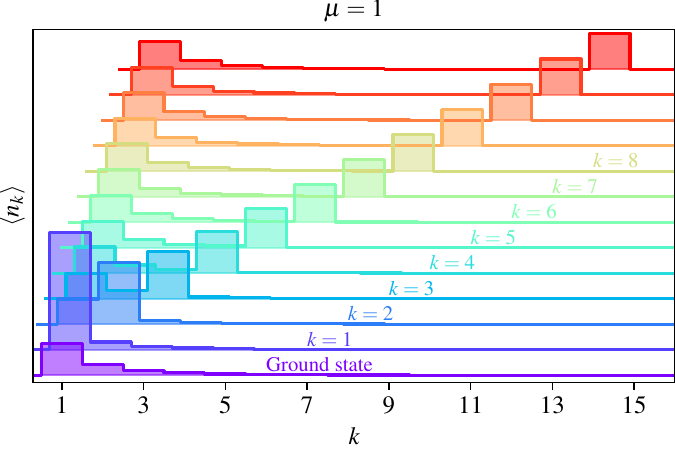}
    \caption{
        Occupation numbers $\langle n_k\rangle$ for the lowest few scar states at strong coupling $\mu=1$. 
  \label{fig:scar_properties}}
\end{figure}

\emph{Tests of the quasiparticle interpretation} -- Let us now test the quasiparticle interpretation of the above found scar-like states in DSG, which admits a quasiparticle description \cite{Mussardo_DSG,Delfino-Mussardo_DSG,Takacs_DSG}. To check if scar-like states follow the relativistic dispersion relation, it is useful to plot the DSG spectra on the $(P,E)$ plane. 
For Dirichlet boundary conditions, as in our simulations, the expectation value of the total momentum $P$ in energy eigenstates vanishes \cite{SM}; however, the momentum content of a state's quasiparticles can be estimated by measuring $P^2$. A one-quasiparticle state, for example, corresponds to a standing wave superposition of a quasiparticle moving to both directions with equal and opposite momenta, and $\langle P^2 \rangle$ is proportional to the square of its momentum. Fig.~\ref{fig:E-P} shows a scatter plot of $\tilde{P}=\sqrt{\langle P^2 \rangle}$, which is an extensive quantity, versus $E$, for the numerical DSG spectrum. The QMBS states are prominently aligned parallel to the `light cone' line $E=\tilde{P} $. Interpreted in view of the previous observations, this means that the scar-like states can be identified as one-quasiparticle states of the above theoretical description. 

To further validate this interpretation, we investigate the characteristics of scar states and compare with theoretical predictions based on the above. Fig.~\ref{fig:scar_properties} shows the occupation numbers $n(k)$ of the CFT harmonic modes computed in the ground state and scar states of DSG at $\mu = 0.5$. The interacting ground state is significantly different from the CFT vacuum, having occupations that perfectly match those of a `squeezed vacuum' state \cite{SM}. Next, we observe that the scar states have approximately the same mode occupations as the ground state except for an additional peak at only one mode each, which is a clear indication that each scar contains a single quasiparticle with different wave numbers. Based on this description of scar states as `squeezed excited states', we can analytically estimate the expectation values of local observables in them. For example, $V(\beta_1)$ is expected to approach a horizontal line for large energies  \cite{SM}, and this is indeed the behaviour of scar states as shown in Fig.~\ref{fig:2}. Note that the thermal line is clearly non-horizontal at large energies and is, instead, expected to approach the CFT thermal line. 

\emph{Deviations from the ETH} -- Several important conclusions can be drawn from the above interpretation. First, the observed QMBS are nothing but one-quasiparticle states of the interacting QFT model we studied and should be present throughout the entire range of the energy spectrum. Second, they are expected to persist for arbitrary interaction strength beyond the values explored numerically. Third, their property of violating the ETH is guaranteed by the fact that they will always be located far from the bulk of other eigenstates and aligned along curves that diverge from the thermal average line. This is true, because:

(1) \emph{the thermal average line of a local observable in any QFT model that is a perturbation of a CFT by a relevant operator, tends to become parallel to that of the CFT in the high energy limit}, and  

(2) \emph{one-quasiparticle states are located along a line determined by relativistic kinematics.} 

Indeed, the entire spectrum of a relativistic QFT is symmetric under boosts in the thermodynamic limit, and any state of the family of one-quasiparticle states can be brought to the bottom edge of the spectrum by such a boost. Therefore, all states in the vicinity of the hyperbolic line should be considered as the “edge” of the spectrum where nonrandom states can be present. Given that for a general observable these two lines diverge, the above statements, which are expected to hold in the thermodynamic limit, imply that the strong version of the ETH cannot hold, whereas the weak version of the ETH may still be valid. Note, however, that, as shown in Fig.~\ref{fig:2}.d, the numerically computed density of states close to the QMBS line does not show any tendency to decrease for increasing $\mu$. At the same time, it is not clear if some generalised version of the ETH could hold as a result of Lorentz symmetry: Unlike other symmetries, Lorentz transformations leave the action rather than the Hamiltonian invariant, and the energy spectrum is constrained by the symmetry in a nontrivial way.

\emph{Discussion} -- Clearly, the above scenario suggests a strong distinction between lattice models and QFT. Perturbing a free lattice model by a generic interaction would not give rise to such atypical eigenstates, because one-quasiparticle states are limited to a finite energy range close to the ground state. On the contrary, in QFT, Lorentz invariance, which is exactly valid in the thermodynamic limit, imposes strong kinematical constraints on the spectrum at all energy scales. Motivated by the above arguments based on RG theory and on Lorentz symmetry, we expect that analogous deviations from the ETH hold in any relativistic nonintegrable QFT that admits a quasiparticle description.

\

\begin{acknowledgments}

We thank Per Moosavi, Alexios Michailidis, Elias Kiritsis and Gabor Takacs for useful discussions.
We thank the support of the HPCaVe computational platform of Sorbonne University, where the main part of the computation has been performed. M.S. is grateful for the environment provided by ISCD and the resources received during his Ph.D. under the supervision of M. Casula and R. Vuilleumier.
S.S. acknowledges support by the European Union’s Horizon 2020 research and innovation program under the Marie Sk\l{}odowska-Curie Grant Agreement No.~101030988.
T.P. acknowledges support from the Program P1-0402 of Slovenian Research Agency (ARRS).
\end{acknowledgments}

\bibliography{Citations.bib}

\newpage
\onecolumngrid
\appendix

\section*{Ergodicity Breaking and Deviation from Eigenstate Thermalisation in Relativistic Quantum Field Theory: \\ Supplemental Material}

%%%
In this Supplemental Material: 
\begin{itemize}
    \item we present the perturbative argument for the presence of scar-like states at weak interaction,
    \item we discuss the technical details on constraints of relativistic kinematics on the QFT spectrum,
    \item we present technical information on the construction of the total momentum operator for Dirichlet boundary conditions,
    \item we derive the observable expectation values in thermal states,
    \item we derive and analyse observable expectation values in one-quasiparticle eigenstates.
\end{itemize}
%%%

\section{Perturbative argument for the presence of scar-like states at weak interaction}
\label{sec:Perturbation}

As mentioned in the main text, the presence of special scar-like states at weak interaction can be explained by perturbation theory, taking into account that the unpertubed model is a CFT and CFTs are highly degenerate models. Here, we elaborate more on the perturbative argument.

The spectrum of a CFT consists of degenerate energy sectors at equidistant energies, with degeneracy numbers increasing like the exponential of the square root of the energy. At leading order in perturbation theory, an interaction mixes states only within the same degenerate sector, resulting in decoupled energy spectra. At next to leading order, the decoupled sectors start mixing with each other but, since the eigenstates are continuous functions of the perturbation strength, some qualitative features originating from this structure may disappear slowly. The decoupled sector spectra may be characterised by chaotic level spacing statistics even at leading perturbative order, as in the DSG \cite{Srdinsek2021}, if the dimensions of the degenerate sectors are large and the interaction behaves like a random matrix (in contrast to the case of non-degenerate perturbation theory, where level crossings persist at first order in perturbation strength\cite{Szasz-Schagrin2021}). Nevertheless, it is expected that states located at the edges of each of the sector spectra are less random, meaning that they do not deviate significantly from the unperturbed states they originate from. In DSG at small $\mu$, the lowest $N$ eigenstates that look like scars in Fig.~1
%\ref{fig:1} 
and Fig.~2 of the main text
%\ref{fig:2} 
are indeed located at one of the two edges of each sector. One way to explain why such eigenstates are less random than others is that the unperturbed basis states they originate from are states with only one particle occupying the highest energy mode in each degenerate sector, which are far more sparsely connected to other basis states in the same sector when the normal-ordered interaction operator is applied on them. 

\section{Relativistic kinematical constraints on the QFT spectrum} 

In this appendix we briefly discuss the constraints on the spectrum of a QFT that admits a quasiparticle description that originate from relativistic kinematics. 

Let us consider the spectrum of a system of {non-interacting} particles in infinite $1D$ space. For a single particle of mass $M$ the energy $E$ and momentum $P$ satisfy the relation $E=\sqrt{P^{2}+M^2}$, therefore on the $(P,E)$ plane the allowed single-particle states form a continuous line (hyperbola) parametrised by the free real parameter $P$. For two free particles there are two independent variables, the particle momenta, so the allowed states span the entire area above the hyperbola $E = 2\sqrt{(P/2)^{2}+M^{2}}$ on the $(P,E)$ plane with $P$ and $E$ now being the total momentum and energy. The kinematics of {non-interacting} multi-particle states is increasingly more complex but otherwise following analogous kinematical constraints. When interactions are present, not much changes {regarding these constraints.} Particles can either form a bound state, which behaves as a distinct composite quasiparticle, or scatter with each other, possibly producing more particles (if their rest energy exceeds the corresponding threshold). Each of these cases corresponds to a distinct family of states on the $(P,E)$ plane, enveloped by a line that approaches the light-cone lines $E=|P|$ at large $P$.  

\begin{figure}
    \centering
  \includegraphics[width = 0.6\linewidth]{{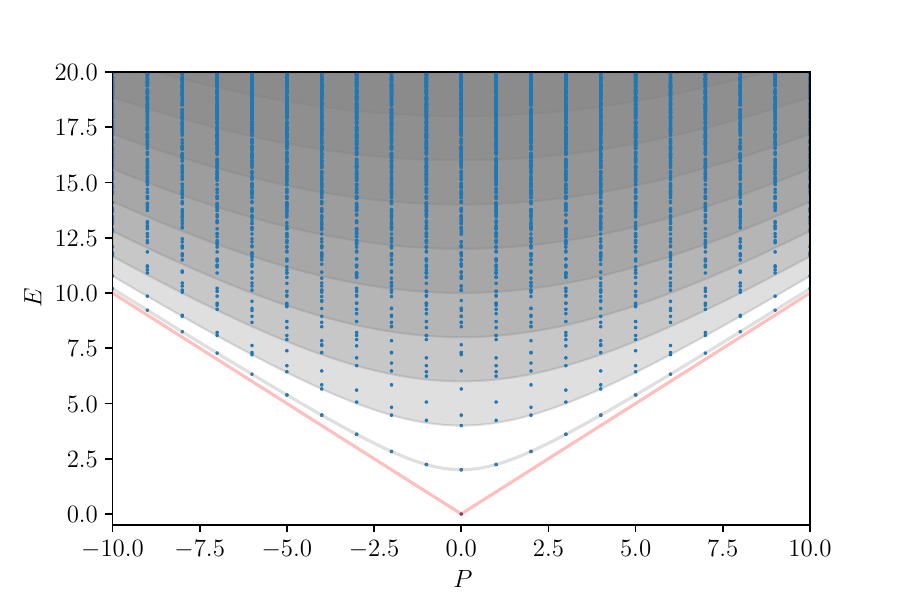}}
    \caption{{
    Illustration of the spectrum of a massive QFT on the $(P,E)$ plane. In infinite space, the spectrum is continuous with the exception of the isolated ground state. Single quasiparticle states form a hyperbola with minimum at $P=0$ equal to the quasiparticle mass, and multi-particle states span continuous areas enveloped by it. The points represent eigenstates of the same QFT in a circle of finite but large size with periodic boundary conditions. 
    \label{fig:E-P0}
  }}
\end{figure}

This description naturally applies also to QFT spectra, where the role of particles is now played by quasiparticles. Signatures of Lorentz invariance are clearly manifested in the structure of their spectra on the $(P,E)$ plane, more specifically, in the hyperbolic lines that envelop each of the different families of eigenstates, as illustrated in Fig.~\ref{fig:E-P0}. In systems of finite size (which is large compared to the inverse of the lightest quasiparticle mass), the spectrum is discrete instead of continuous but is otherwise constrained by the same kinematical rules, up to finite size corrections to the quasiparticle parameters \cite{Luscher1986i,Luscher1986ii}. 

%%%%%%%%

\section{Construction of the total momentum operator for Dirichlet boundary conditions \label{app:P}}

In the bosonic QFT models studied here, the total momentum operator $P$ is defined as 
\begin{equation}
    P=-\int_{0}^{L}\mathrm{d}x\,\Pi(x)\partial_{x}\phi(x),
\end{equation}
Substituting the mode expansions of the field operator $\phi$ and the canonically conjugate field $\Pi$, which for Dirichlet boundary conditions are
\begin{align}
\phi(x) & =\sqrt{\frac{2}{\pi}}\sum_{n=1}^\infty\frac{1}{\sqrt{2n}}(a_{n}+a_{n}^{\dagger})\sin(n\pi x/L) \label{app:phi} \\
\Pi(x) & =-\mathrm{i}\frac{\pi}{L}\sqrt{\frac{2}{\pi}}\sum_{n=1}^\infty\sqrt{\frac{n}{2}}(a_{n}-a_{n}^{\dagger})\sin(n\pi x/L) \label{app:pi}
\end{align}
we obtain 
\begin{align}
P & = \mathrm{i} (\pi/L)\sum_{n_{1},n_{2}}\sqrt{n_{1}n_{2}}F(n_{1},n_{2})(a_{n_{1}}-a_{n_{1}}^{\dagger})(a_{n_{2}}+a_{n_{2}}^{\dagger}) \label{app:P}
\end{align}
where
\begin{align*}  
    F(n_{1},n_{2}) & := \frac1L \int_{0}^{L}\mathrm{d}x\,\sin(n_{1}\pi x/L)\cos(n_{2}\pi x/L) 
    =\begin{cases}
(2/\pi)\frac{n_{1}}{n_{1}^{2}-n_{2}^{2}} & \text{if }n_{1}+n_{2}\text{ odd}\\
0 & \text{if }n_{1}+n_{2}\text{ even}
\end{cases}
\end{align*}
Note that $F(n,n)=0$ so the above expression for $P$ is normal-ordered. 

Unlike for periodic boundary conditions, in the present case the total momentum operator $P$ is not diagonal in the energy eigenbasis since the Dirichlet boundary conditions break translational invariance, therefore the commutator $[H,P]$ is not zero. 
%\new{In fact, the naive definition of $P$ is not even self-adjoint.}
% add \cite{ALBRECHT2023}

The momentum operator $P$ anticommutes with parity $S$ (spatial reflection with respect to the middle)
\begin{equation}
    SP=-PS    
\end{equation}
The basis states $|\Psi\rangle$ are eigenstates of $S$ with eigenvalues $s_\Psi = \pm1$ since $S^2=1$. The matrix elements of $P$ between two basis states $|\Psi\rangle,|\Psi'\rangle$ with the same parity vanish, because 
\begin{align*}
    \left\langle \Psi|P|\Psi'\right\rangle & =\left\langle \Psi|PS^{2}|\Psi'\right\rangle \\
	& =-\left\langle \Psi|SPS|\Psi'\right\rangle \\
	& =-s_{\Psi}s_{\Psi'}\left\langle \Psi|P|\Psi'\right\rangle 
\end{align*}
i.e. if their parity eigenvalues are equal, $s_{\Psi}=s_{\Psi'}$, then $\left\langle \Psi|P|\Psi'\right\rangle =0$. Therefore $P$ is a block off-diagonal matrix with non-vanishing matrix elements only between opposite parity sectors.  Moreover, since the Hamiltonian commutes with parity
\begin{equation}
    SH=HS    
\end{equation}
the energy eigenstates are also eigenstates of $S$. 

From the above we conclude that the expectation values of $P$ in energy eigenstates vanish. This is why to obtain an estimate of the quasiparticle momentum content of the energy eigenstates $|E\rangle$ we compute $\sqrt{\langle E| P^2 |E \rangle} $. \new{For further details about the $P$ operator under Dirichlet boundary conditions, see \cite{Albrecht_2023}.}

\section{Observable expectation values in thermal states}

Thermal expectation values of local or extensive observables at large temperatures converge to those of the unperturbed CFT model, therefore it is useful to calculate the latter exactly, either analytically or numerically, for comparison with the approximate numerical data for the perturbed model. In particular, we focus on the number operator $N$. 

The expectation value of $N$ in a thermal state of the free massless boson Hamiltonian at temperature $T$ is
\begin{equation}
    \langle N \rangle_{\new{T}} = \sum_k \frac{1}{\exp(\omega_k/T)-1},
\end{equation}
where $\omega_k = \pi k/L$, with $k=1,2,3,\dots$ are the frequencies of the harmonic sine modes. The series can be evaluated analytically and equals 
\begin{equation}
    \sum_{n=1}^\infty \frac{1}{\exp(n \omega/T)-1} = 
-(T/\omega) \psi _{e^{\frac{\omega}{T}}}^{(0)}(1)-(T/\omega) \log
   \left(e^{\frac{\omega}{T}}-1\right)+\frac{1}{2},
\end{equation}
where the function $\psi_{e^{\omega/T}}^{(0)}(x)$ is the digamma function. 

\section{Observable expectation values in one-quasiparticle eigenstates}

As explained in the main text, the scar-like states observed in the numerical spectra can be interpreted as one-quasiparticle states. A quasiparticle of a nonintegrable interacting model like the DSG is a collective excitation above the ground state, which is generally represented by highly nonlinear and nonlocal ladder operators in any noninteracting basis and their parameters depend nonperturbatively on the interaction. Despite this, eigenstates corresponding to one massive quasiparticle are relatively simple in that no effects of quasiparticle scattering or production are present. For this reason their characteristics are fully determined by the relativistic kinematics of a single particle with all nontrivial effects of strong interaction concealed in the particle's self-energy $\Sigma$. The latter corresponds to a momentum dependent shift of the quasiparticle's effective mass and at large momenta it asymptotically tends to a constant value. Observables in such eigenstates can therefore be calculated using the Bogoliubov transformation that relates the creation and annihilation operators of the original free massless particle to a free massive quasiparticle. 

% Expanding the field $\phi$ in Fourier modes and ladder operators for each mode we have 
% \[
% \phi = \sum_{k=1}^\infty \frac{1}{\sqrt{ 2 k \pi }}\sin(k \pi x/L) (a_k + a_{k}^\dagger)
% \]
%%%%%
% from which we find
% \[
% H_0 = \frac{\pi}{L} \sum_k k a_k^\dagger a_k % or: (a_k^\dagger a_k +1/2) if not normal-ordered.
% \]
% and 
% \[
% V(\beta) = \int_0^L \prod_{k=1}^\infty e^{i\beta \frac{1}{\sqrt{ 2 k \pi }}\sin(k \pi x/L) a_k^\dagger } e^{i\beta \frac{1}{\sqrt{ 2 k \pi }}\sin(k \pi x/L) a_k} \; \mathrm{d}x + \mathrm{h.c} 
% \]
% \[
% = (L/\pi) \int_0^\pi \prod_{k=1}^\infty e^{i\beta \frac{1}{\sqrt{ 2 k \pi }}\sin(k s) a_k^\dagger } e^{i\beta \frac{1}{\sqrt{ 2 k \pi }}\sin(k s) a_k} \; \mathrm{d}s + \mathrm{h.c}
% \]
% from which we see that $H_0 \propto 1/L$ and $V(\beta) \propto L $. 
%%%%%

\new{We will consider the interaction operator $V(\beta)$ as an observable. As mentioned in the main text, this operator is defined as the space integral of the normal-ordered cosine operator, which is the hermitian part of the operator $\mathcal{V}_{\beta}(x)$ defined as} 
\begin{align}
    \mathcal{V}_{\beta}(x) & =\,:\!\mathrm{e}^{\mathrm{i}\beta\phi(x)}\!:_{0}  \nonumber \\
    & = \prod_{k=1}^\infty \,:\!\exp\left[{\mathrm{i}\beta{\frac{1}{\sqrt{k\pi}}}(a_{k}+a_{k}^{\dagger})\sin(k\pi x/L)}\right]\!:_{0}\, \nonumber \\
    & = \prod_{k=1}^\infty \exp\left[{\mathrm{i}\beta{\frac{1}{\sqrt{k\pi}}} a_{k}^{\dagger} \sin(k\pi x/L)}\right] \, \exp\left[{\mathrm{i}\beta{\frac{1}{\sqrt{k\pi}}} a_{k} \sin(k\pi x/L)}\right] \label{eq:vertex}
\end{align}
where normal-ordering refers to the free massless boson basis with Dirichlet boundary conditions \new{and in the second line we used the expansion (\ref{app:phi})}.

\new{We note in passing that, since $\mathcal{V}_{\beta}(x)$ is a function of $x/L$ only (instead of $x$ and $L$ separately), the spatially integrated operator $V(\beta)$ is a linear function of $L$ 
\begin{align}
    V(\beta) = \int_0^L\mathrm{\mathrel{d}}x\,\mathcal{V}_{\beta}(x) 
    = \frac{L}{\pi} \int_0^\pi \mathrm{\mathrel{d}}s \, \prod_{k=1}^\infty \,:\!\exp\left[{\mathrm{i}\beta{\frac{1}{\sqrt{k\pi}}}(a_{k}+a_{k}^{\dagger})\sin(k s)}\right]\!:_{0}\,
\end{align}
On the contrary, the free Hamiltonian $H_0$ is inversely proportional to $L$ 
\begin{align}
H_0 = \frac{\pi}{L} \sum_{k=1}^\infty k (a_k^\dagger a_k + \tfrac{1}{2}) \label{qpp:H0} 
\end{align}
The same holds also for the total momentum operator $P$ shown in (\ref{app:P}). 
% } 
% \new{
Note that in TCSA studies \cite{TCSA-sG1,Takacs_DSG} it is customary to use the CFT normalisation of `vertex' operators, which differs from the normal-ordered operator (\ref{eq:vertex}) by an additional factor 
\[
\mathcal{\tilde{V}}_\beta(x) \equiv (2\pi/L)^{\beta^2/(4\pi)} :\! \mathrm{e}^{\mathrm{i}\beta \phi} \!:_0 \; = (2\pi/L)^{\beta^2/(4\pi)} \mathcal{{V}}_\beta(x)
\] 
However, constructing the DSG Hamiltonian based on this choice, we find that the two cosine perturbations scale differently in the limit $L\to\infty$ due to their different frequencies, therefore the model approaches the SG limit which is integrable \cite{Bajnok_2001}. 
The level spacing statistics of the spectrum then stops following RMT predictions and becomes Poissonian \cite{Srdinsek2021}. 
In order to stay away from the integrable limits for any value of $L$, we use the $L$ independent normalisation of (\ref{eq:vertex}) and choose the mixing parameters $\lambda_1$ and $\lambda_2$ equal, as specified in the main text.}  

% \new{Using the mode expansions (\ref{app:phi}) and (\ref{app:pi}) 
% \begin{align}
% H_0 = \frac{\pi}{L} \sum_k k a_k^\dagger a_k
% \end{align}
% }

%(alternatively, if one uses the CFT normalisation $\mathcal{V}_\beta(x)$ then the corresponding coefficients $\lambda'_i=\lambda_i (2\pi/L)^{\beta_i^2/(4\pi)}$ with which they enter in the DSG Hamiltonian must be rescaled with $L$ so as to cancel the extra $L^{-\beta_i^2/(4\pi)}$ factors).} 

\new{The matrix elements of $\mathcal{V}_{\beta}(x)$} in the massless boson basis states $|\Psi_{0}\rangle$, which are characterised by the occupation numbers $\{n_{0}(k)\}$ in each mode $k$, are
\[
    \left\langle \Psi'_{0}\left|\mathcal{V}_{\beta}(x)\right|\Psi_{0}\right\rangle =\prod_{k=1}^{\infty}F_{n_{0}'(k),n_{0}(k)}\left(\beta;k,x\right)
\]
with \cite{Horvath2022}
\begin{align*}
    F_{n',n}(\beta;k,x) & := 
    \left\langle n'\left| \mathrm{e}^{\mathrm{i}\beta\sqrt{\frac{1}{k\pi}} a^{\dagger} \sin(k\pi x/L)} \, \mathrm{e}^{\mathrm{i}\beta\sqrt{\frac{1}{k\pi}} a \sin(k\pi x/L)} \right|n\right\rangle \\ 
    & =\sum_{j=\text{max}(0,n-n')}^{n}
 \left(\mathrm{\mathrel{i}}\frac{\beta}{\sqrt{4\pi k}}\sin(k\pi x/L)\right)^{2j+n'-n}\frac{(-1)^{j+n'-n}\sqrt{n!n'!}}{j!(n-j)!(j+n'-n)!}
\end{align*}

To compute the \new{expectation values of $\mathcal{V}_{\beta}(x)$} in a general eigenstate $|\Psi_{M}\rangle $ of a free massive particle basis state $|\Psi_{M}\rangle$, characterised by the corresponding occupation numbers $\{n_{M}(k)\}$ corresponding to particle mass $M$, we use a Bogoliubov transformation to switch from the massless to the massive basis
\begin{align*}
    b_k & = \mu(k) a_k + \nu(k) a_k^\dagger
\end{align*}
with 
\begin{align*}
    \mu(k) & = \frac{1}{2} \left( \sqrt{\frac{E_M(k)}{E_0(k)}} + \sqrt{\frac{E_{\new{0}}(k)}{E_{\new{M}}(k)}} \right) \\ 
    \nu(k) & = \frac{1}{2} \left( \sqrt{\frac{E_M(k)}{E_0(k)}} - \sqrt{\frac{E_{\new{0}}(k)}{E_{\new{M}}(k)}} \right) 
\end{align*}
Using the Baker-Campbell-Hausdorff formula to find the relation between the \new{operator $\mathcal{V}_{\beta}(x)$} normal-ordered in the free massless and massive bases, we obtain the general formula for matrix elements of $\mathcal{V}_\beta(x)$ in the free massive basis 
\begin{equation}
\left\langle \Psi'_{M}\left|\mathcal{V}_{\beta}(x)\right|\Psi_{M}\right\rangle =\prod_{k=1}^{\infty}\mathrm{e}^{-\tfrac{1}{2}\frac{\beta^{2}}{4\pi k}\left(\lambda_{M}(k)-1\right)}F_{n_{M}'(k),n_{M}(k)}\left(\beta\lambda_{M}(k);k,x\right)\label{app:eq:vertex_matrix_elements_massive}
\end{equation}
where 
\[
\lambda_{M}(k)=\frac{E_{0}(k)}{E_{M}(k)}=\frac{k}{\sqrt{k^{2}+(ML/\pi)^{2}}}
\]
are the ratios of massless to massive mode energies. 

The free massive ground state has all occupation numbers $\{n_{M}(k)\}$
equal to zero and one particle states have all zero except at one
$k$ where $n_{M}(k)=1$. Their energy difference from the ground
state is 
\begin{equation}
E_{M}(k)=\sqrt{(\pi k/L)^{2}+M^{2}},\quad k=1,2,\dots\label{app:eq:massive_mode_energy}
\end{equation}
We denote the free massive ground state as $|\Omega_{M}\rangle$ and
one particle states as $|k;M\rangle$. Being special cases of (\ref{app:eq:vertex_matrix_elements_massive}),
the expectation values of \new{$\mathcal{V}_{\beta}(x)$} in such states are
\begin{align*}
\left\langle \Omega_{M}\left|\mathcal{V}_{\beta}(x)\right|\Omega_{M}\right\rangle  & =1\\
\left\langle k;M\left|\mathcal{V}_{\beta}(x)\right|k;M\right\rangle  & =\mathrm{e}^{-\tfrac{1}{2}\frac{\beta^{2}}{4\pi k}\left(\lambda_{M}(k)-1\right)}F_{1,1}\left(\beta\mu(k),k\right)\\
 & =\mathrm{e}^{-\tfrac{1}{2}\frac{\beta^{2}}{4\pi k}\left(\lambda_{M}(k)-1\right)}\left(1-\frac{\beta^{2}}{4\pi}\frac{\sin^{2}(k\pi x/L)}{\sqrt{k^{2}+(ML/\pi)^{2}}}\right)
\end{align*}
For the spatially integrated %\new{\sout{vertex}} 
operator $V(\beta) = \int\mathrm{\mathrel{d}}x\,\mathcal{V}_{\beta}(x)$, we find
\begin{align*}
\left\langle k;M\left| 
V(\beta) \right|k;M\right\rangle  & =L\mathrm{e}^{-\tfrac{1}{2}\frac{\beta^{2}}{4\pi}\left(\frac{1}{\sqrt{k^{2}+(ML/\pi)^{2}}}-\frac{1}{k}\right)}\left(1-\frac{1}{2}\frac{\beta^{2}}{4\pi}\frac{1}{\sqrt{k^{2}+(ML/\pi)^{2}}}\right)
\end{align*}

We can now use this formula to compute a theoretical prediction for the curve where the quantum scar points are located in the ETH scatter plot for the integrated cosine operator. If we set the length unit to $L=\pi$ for simplicity and express the above expectation value as a function of the state energy $E$ measured from the ground state level as given by (\ref{app:eq:massive_mode_energy}), we find 
\begin{align}
\frac{\left\langle V(\beta) \right\rangle _{\text{1qp}}}{\left\langle V(\beta)
 \right\rangle _{\text{gs}}} & =\mathrm{e}^{-\tfrac{1}{2}\frac{\beta^{2}}{4\pi}\left(\frac{1}{E}-\frac{1}{\sqrt{E^{2}-M^{2}}}\right)}\left(1-\frac{1}{2}\frac{\beta^{2}}{4\pi}\frac{1}{E}\right),\quad E\geq\sqrt{1+M^{2}}
 \label{eq:V_op_EV}
\end{align}
that has one unknown parameter $M$ that can be estimated from the
numerical data by fitting. 
For large $E$ we obtain the simpler relation
\begin{align*}
\frac{\left\langle V(\beta) \right\rangle _{\text{1qp}}}{\left\langle V(\beta)
 \right\rangle _{\text{gs}}} & \underset{E\to\infty}{\sim}1-\frac{1}{2}\frac{\beta^{2}}{4\pi}\frac{1}{E}+\mathcal{O}(E^{-3})
\end{align*}
which is free of unknown parameters. If the quasiparticle interpretation of scar states is correct, these relations suggest that the expectation value of $V(\beta)$ in these states should tend to a horizontal line with $1/E$ corrections at large $E$, which is indeed what we see in Fig.~2 of the main text.

We can similarly calculate the occupation numbers $n(k)$ of the original massless particle modes in the ground state and in excited states of the interacting model with one quasiparticle at momentum $q$, in the same approximation
\begin{align}
\langle n(k)\rangle_{\text{gs}}
\approx \nu(k)^2 = \frac{1}{4} \left( \frac{E_M(k)}{E_0(k)} + \frac{E_0(k)}{E_M(k)} - 2\right) 
\label{eqq:n(k)gs}
\end{align}
which decreases like \new{$1/k^4$} at large $k$, and 
\begin{align}
\langle n(k)\rangle_{\text{1qp}} = \langle q; M | n(k) | q;M \rangle
\approx \langle n(k)\rangle_{\text{gs}} + f(q) \delta_{k,q} 
\label{eqq:n(k)1qp}
\end{align}
where $f(q) = 1+2\nu(q)^2 $, which tends to 1 for large $q$. 

\begin{figure}
\centering
        \includegraphics[width = 0.75\linewidth]{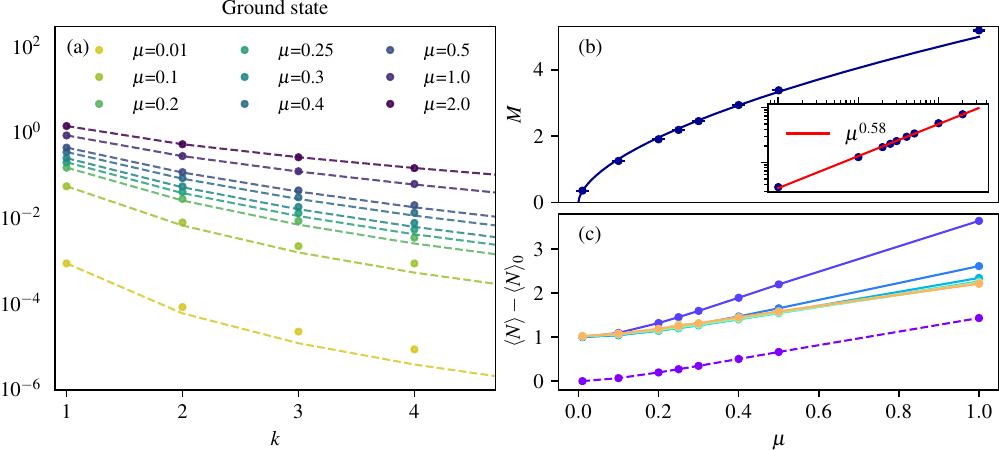} 
    \caption{
        \textit{Properties of the ground state and one-quasiparticle states.} (a) Occupation numbers of a ground state as a function of $\mu$ (\textit{points}) in a log scale compared with a distribution predicted from a massive free boson calculation. Fits are performed by varying the mass parameter. (b) Effective mass $M$ extracted from (a) as a function of interaction strength $\mu$. The inset displays the same data in log-log scale {showing that the dependence is nonanalytic (algebraic with a fractional exponent).}
        (c) Total occupation number of a first few states as a function of $\mu$.
  \label{fig:scar_properties_2}}
\end{figure}

The agreement of the numerical data with the above expressions is shown on Fig.~4 of the main text and Fig.~\ref{fig:scar_properties_2} here. Eq.~(\ref{eqq:n(k)gs}) describes correctly the numerically computed occupation numbers in the ground state of DSG for all values of $\mu$ we used in our simulations. Fitting the numerical data to this formula we can estimate the quasiparticle mass $M$. Plotting it as a function of $\mu$ we verify that its dependence is nonperturbative, since it increases like a noninteger power for small $\mu$. Moreover, comparing numerical data for $\langle n(k)\rangle$ in scar states to Eq.~(\ref{eqq:n(k)1qp}), we verify their consistency which confirms the quasiparticle interpretation of these states. 

\end{document}